\documentclass[preprint,superscriptaddress,showpacs,tightenlines]{revtex4}
\usepackage{amssymb}
\usepackage{amsmath}
\usepackage{graphicx,bm}

\input{tcilatex}

\begin{document}

\title{Conductance of a STM contact on the surface of a thin film }
\author{N.V. Khotkevych}
\affiliation{B.I. Verkin Institute for Low Temperature Physics and Engineering, National
Academy of Sciences of Ukraine, 47, Lenin Ave., 61103, Kharkov, Ukraine.}
\author{Yu.A. Kolesnichenko}
\affiliation{B.I. Verkin Institute for Low Temperature Physics and Engineering, National
Academy of Sciences of Ukraine, 47, Lenin Ave., 61103, Kharkov, Ukraine.}
\author{J.M. van Ruitenbeek}
\affiliation{Kamerlingh Onnes Laboratorium, Universiteit Leiden, Postbus 9504, 2300
Leiden, The Netherlands.}

\begin{abstract}
The conductance of a contact, having a radius smaller than the Fermi wave
length, on the surface of a thin metal film is investigated theoretically.
It is shown that quantization of the electron energy spectrum in the film
leads to a step-like dependence of differential conductance $G\left(
V\right) $ as a function of applied bias $eV$. The distance between
neighboring steps in $eV$ equals the energy level spacing due to size
quantization. We demonstrate that a study of $G\left( V\right) $ for both
signs of the voltage maps the spectrum of energy levels above and below
Fermi surface in scanning tunneling experiments.
\end{abstract}

\keywords{STM, electron tunneling, buried structures, surface states, thin
films.}
\pacs{74.55.+v, 85.30.Hi, 73.50.-h}
\maketitle

\section{Introduction}

Today a fairly large number of papers have addressed the problem of
calculating point-contact conductance for use in analyzing and interpreting
scanning tunneling microscopy (STM) experiments (for reviews see, for
example, \cite{Hofer,Blanco}). The low symmetry of the problem and the wide
variety of objects under study do not allow developing a general theory of
STM, and different approaches for specific problems are used. The theory
papers on this subject can be divided into two groups: One uses methods
taking into account the specific atomic structure of the STM tip and that of
the test specimen. These methods make it possible to reproduce the
crystallographic structure of the sample surface in the calculated STM
images and this is very useful for arriving at a correct interpretation of
experimental data. The main deficiency of this approach is the lack of
analytical formulas for the STM current-voltage characteristics as numerical
calculations must be performed for every specific case. The other group of
works exploit simplified models of noninteracting electrons which allows
finding relatively simple analytical expressions that describe the STM
current qualitatively. For this reason such theoretical results are widely
used by experimentalists.

One of the first free-electron models describing STM experiments was
proposed by Tersoff and Hamann \cite{Tersoff} whose theoretical analysis of
tunnel current is based on Bardeen's formalism \cite{Bardeen}, in which a
tunneling matrix element is expressed by means of independent wave functions
for the tip and the sample within the barrier region. Using the model wave
functions the authors \cite{Tersoff} showed that the conductance of the
system is proportional to the local density of states of the sample at the
tip position. In principle it is possible to extract information on
subsurface objects (single defects, clusters, interfaces etc.) by STM, but
this requires a more detailed theoretical analysis \cite{Kobayashi}, which
takes into account the influence of subsurface electron scattering on the
tunneling current.

The physical picture of an electron tunneling through a classically
forbidden region is that the electron flow emerging from the barrier is
defined by the matching of the wave functions of carriers incident on the
barrier and those that are transmitted. For a three dimensional STM geometry
the wave functions for electrons transmitted through the vacuum region
radically differs from the electron wave functions in an isolated sample and
they describe the electron propagation into the bulk from a small region on
the surface below the STM tip. In contrast, the theory of Ref.~\cite{Tersoff}
and its modifications (see \ \cite{Hofer,Blanco} and references therein)
uses unperturbed wave functions of the surface Bloch states. Changes in the
wave functions of transmitted electrons due to scattering by subsurface
objects provide the information about such scattering in the STM conductance.

In Ref.~\cite{Avotina2005} it was proposed to introduce in the theory of STM
the model by Kulik \textit{et al.} \cite{Kulik}. In this model a three
dimensional STM tip is replaced by an inhomogeneous barrier in an otherwise
nonconducting interface that separates the two conductors. In Ref.~\cite%
{Kulik} it was shown that under assumption of small transparency of the
tunnel barrier the wave function (and thus the current-voltage
characteristics) can be found analytically for an arbitrary size of the
tunnel area. The results in \cite{Avotina2005} for the conductance of the
tunnel point-contact were generalized to an arbitrary Fermi surface for the
charge carriers in Ref.~\cite{Avotina06}. In a series of papers the model 
\cite{Kulik} has been expanded to describe oscillations of the STM
conductance resulting from electron scattering by subsurface defects (for
reviews see \cite{AKR,Khotkevych}).

Scanning tunneling microscopes have been widely used for the study of
various small-sized objects: islands, thin films deposited on bulk
substrates, etc \cite{Altfeder1997}-\cite{Gasparov}. First, a discrete
periodic spatial variation of the STM current originating from the
quantization of electron states was observed in the quantum wedge: a
nanoscale flat-top Pb island on a stepped Si(111) surface \cite{Altfeder1997}%
. Later these authors showed that that the lattice structure of an interface
buried under a film of Pb, whose thickness can be as many as 10 times the
Fermi wavelength, can be clearly imaged with STM \cite{Alt2}. They concluded
that the key to the transparency of a metal lies in a highly anisotropic
motion of the electrons and the strong quantization of their transverse wave
function components. In the paper \cite{Jiang} the electronic states of thin
Ag films grown on GaAs(110) surfaces was investigated by STM with
single-layer thickness resolution, and the quantum-well states arising from
the confinement geometry of the Ag films have been identified. Quantum size
effects, manifested in the formation of new electronic bound states, were
investigated by STM on thin Pb islands of varying heights on the Si(111)-(7$%
\times $7) surface in Ref.\cite{Su}. In experiments \cite{Alt} it was
demonstrated that scanning tunneling microscopy and spectroscopy of
epitaxial Pb islands on Si(111) reveal adiabatic lateral modulation of the
energy spectra of the quantum well, providing remote electronic images of
the subsurface reflection phase. In Ref.~\cite{Hongbin} a step structure at
the buried Pb on Si(111) 6$\times $6-Au interface was determined by
utilizing the presence of quantum well states. It was demonstrated that the
spatial step positions as well as the step heights can be extracted
nondestructively and with atomic layer precision by STM. Vertical Friedel
oscillations in interface-induced surface charge modulations of Pb islands
of a few atomic layers on the incommensurate Si(111)-Pb surface have been
observed \cite{Jian}. Thus, detailed experimental results have been
obtained, but a microscopic theory for STM tunneling spectra on samples of
finite size has not been reported, which provides the motivation for the
present work. Current-voltage characteristics for size quantization in
planar thin film geometries of metal-insulator-metal tunneling junctions
have been investigated theoretically in Refs.~\cite{Davis1,Khachaturov}.
Standing electron wave states in thin Pb films have been observed by
electron tunneling in early experiments by Lutskii \textit{et al.} \cite%
{Lutskii}.

In this paper we present the differential conductance $G\left( V\right) $
for small contacts, having a radius $a$ smaller than the Fermi wave length $%
\lambdabar _{\mathrm{F}}=\hbar /p_{\mathrm{F}},$ where $p_{\mathrm{F}}$ is
the Fermi momentum. The contacts are formed on the surface of a thin metal
film and we analyze the voltage dependence of $I\left( V\right) $\ and%
\textbf{\ }$G\left( V\right) $. We focus on the size quantization effects of
the electron energy spectrum in the film on $G\left( V\right) $.

The organization of this paper is as follows. The model that we use to
describe the contact, and the method for obtaining a solution of the
three-dimensional Schr\"{o}dinger equation asymptotic in the small radius of
the contact, are described in Sec.~II. In Sec.~III the current-voltage
characteristics and the differential conductance are found on the basis of a
calculation of the probability current density through the contact. Sec.~IV
presents a physical interpretation of the results obtained. In Sec.~V we
conclude by discussing the possibilities for exploiting these theoretical
results for interpretation of electron energy spectroscopy in thin films by
STM. In the appendixes we solve the Schr\"{o}dinger equation for the tunnel
point contact in framework of our model (App.I) and for a point contact
without barrier (App.II) and find the wave functions for electrons
transmitted through the contact. These solutions are used in Sec.III for the
calculation of current.

\section{Model and electron wave function of the system}

The model that we consider is illustrated in Fig.~1. Electrons can tunnel
through an orifice centered at the point $\mathbf{r}=0$ in an infinitely
thin insulating interface at $z=0$ from a conducting half-space (the tip)
into a conducting sheet of thickness $d$ (Fig.1(b)). The radius $a$ of the
contact and the thickness $d$ of the film are assumed to be much smaller
than the shortest mean free path, i.e we consider a purely ballistic
problem. 
\begin{figure}[tbp]
\includegraphics[width=15cm,angle=0]{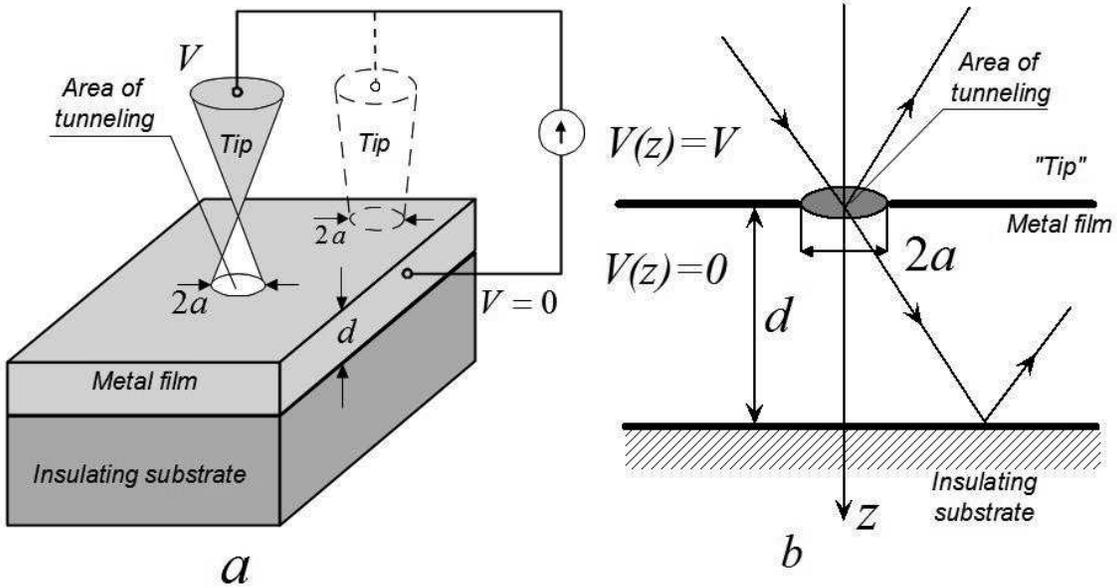}
\caption{Schematic representation of a STM experiment on a thin metal film
(a) and the model that we employ to represent the contact between a bulk
conductor (tip) and a metallic film (b). The dashed picture of the tip in
(a) illustrates a metallic point-contact (STM tip touches the surface).
Electron trajectories in (b) are shown schematically. }
\label{fig1}
\end{figure}
The wave function $\psi $ satisfies the Schr\"{o}dinger equation \ \ \ 
\begin{equation}
\triangledown ^{2}\psi \left( \mathbf{r}\right) +\frac{2m^{\ast }}{\hslash
^{2}}\left[ \varepsilon -U\left( \mathbf{r}\right) \right] \psi \left( 
\mathbf{r}\right) =0.  \label{Schrod}
\end{equation}%
In Eq.(\ref{Schrod}) $\ m^{\ast }$ and $\varepsilon $ are electron effective
mass and energy, respectively. The inhomogeneous potential barrier in the
plane $z=0$ we describe by the function $U\left( \mathbf{r}\right)
=U_{0}f\left( \mathbf{\rho }\right) \delta \left( z\right) $, where $\mathbf{%
\rho }=\left( x,y\right) $ is a two dimensional position vector in the plane
and $f^{-1}\left( \mathbf{\rho }\right) =\Theta \left( a-\rho \right) $,
with $\Theta \left( x\right) $ the Heaviside step function. For such model
the wave function $\psi \left( \mathbf{r}\right) $ satisfies the following
boundary conditions at the interface $z=0$ and at the metal sheet surface $%
z=d$ 
\begin{equation}
\psi \left( \mathbf{\rho },+0\right) =\psi \left( \mathbf{\rho },-0\right) ,
\label{z=0}
\end{equation}%
\begin{equation}
\psi _{z}^{\prime }\left( \mathbf{\rho },+0\right) -\psi _{z}^{^{\prime
}}\left( \mathbf{\rho },-0\right) =\frac{2m^{\ast }U_{0}}{\hbar ^{2}}f\left( 
\mathbf{\rho }\right) \psi \left( \mathbf{\rho },0\right) ,  \label{jump}
\end{equation}%
\begin{equation}
\psi \left( \mathbf{\rho },d\right) =0.  \label{z=d}
\end{equation}%
Eqs.~(\ref{Schrod})-(\ref{z=d}) can be solved in the limit of a small
contact, $ka\ll 1$ ($k=\sqrt{2m^{\ast }\varepsilon }/\hslash $ is the
absolute value of the electron wave vector $\mathbf{k}$). In the zeroth
approximation in the contact diameter the solutions of Eq.(\ref{Schrod}) for 
$z\gtrless 0$ are independent and satisfy the zero boundary condition $\psi (%
\mathbf{\rho },0)=0$ at the impenetrable interface at $z=0.$ The quantum
states in the conducting half-space $\left( z<0\right) $ (the tip) are
defined by the three components of the electron wave vector $\mathbf{k}%
=\left( \mathbf{k}_{\parallel },k_{z}\right) $, with $\mathbf{k}_{\parallel }
$ a two-dimensional vector parallel to the interface. In the metal film $%
\left( 0<z<d\right) $ the quantum states are characterized by a
two-dimensional vector \textbf{$\kappa $} perpendicular to the $z-$axis and
by the discrete quantum number $n$ $\left( n=1,2,...\right) $ which results
from the finite size of the conductor in the $z$ direction. The energy
eigenvalues and eigenfunctions for the two disconnected conductors are given
by 
\begin{gather}
\varepsilon =\frac{\hbar ^{2}\left( k_{\parallel }^{2}+k_{z}^{2}\right) }{%
2m^{\ast }}\equiv \frac{\hbar ^{2}k^{2}}{2m^{\ast }},  \label{epsz<0} \\
\psi _{0}\left( \mathbf{r}\right) =2ie^{i\mathbf{k}_{\parallel }\mathbf{\rho 
}}\sin k_{z}z,\qquad z<0;  \label{psi0z<0}
\end{gather}%
and%
\begin{eqnarray}
\varepsilon  &=&\frac{\hbar ^{2}\left( \kappa ^{2}+k_{zn}^{2}\right) }{%
2m^{\ast }};\quad n=1,2,...,  \label{epsz>0} \\
\psi _{0}\left( \mathbf{r}\right)  &=&-2ie^{i\mathbf{\kappa \rho }}\sin
k_{zn}z,\qquad 0<z<d,  \label{psi0z>0}
\end{eqnarray}%
where $k_{zn}=\pi n/d.$ In Eqs.~(\ref{psi0z<0}) and (\ref{psi0z>0}) we use a
wave function normalization with unit amplitude of the wave incident to the
interface.

The partial wave for the first order approximation $\psi _{1}\left( \mathbf{r%
}\right) $ in the small parameter $ka\ll 1$, which describes the transition
of electrons from one to the other conductor, is given in the appendices.
Appendix I, Eqs.~(AI.5) and (AI.6), gives the solution for a tunnel point
contact, having a potential barrier of small transparency $t=k\hslash
^{2}/m^{\ast }U_{0}\ll 1$ at the orifice in the plane $z=0$. Appendix II,
Eqs. (AII.6) - (AII.8), gives the solutions for a contact without barrier.
Figure 2 illustrates the spacial variation of the square modulus of the wave
function for electrons transmitted through the contact into the film. 
\begin{figure}[tbp]
\includegraphics[width=10cm,angle=0]{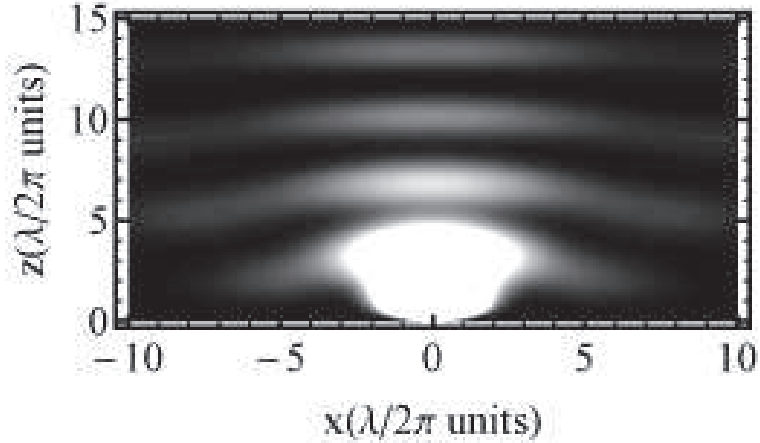}
\caption{Space distribution of the square modulus of the wave function for
electrons injected by a STM tip into a metal sheet of thickness $%
d=15\lambdabar ,$ where $\protect\lambda =2\protect\pi /k$ is the electron
wave length, $\lambdabar =\protect\lambda /2\protect\pi $.}
\label{wf}
\end{figure}

\section{Current-voltage characteristic and conductance of a point contact.}

As has been shown in Ref.~\cite{KOSh} for a ballistic point contact of small
radius $a,$ with $a$ much smaller than the electron mean free path $l$, the
electrical potential $V(\mathbf{r})$ drops over a distance $r\sim a$ from
the contact, and in the limit $a\rightarrow 0$ the potential $V(\mathbf{r})$
can be approximated by a step function $V\,\Theta (-z)$. In this
approximation, for the calculation of the electrical current we can take the
electron distribution functions $f^{\left( \mp \right) }$ at $z\lessgtr $ $0$
as the Fermi functions $f_{\mathrm{F}}$ with energies shifted by the applied
bias $eV$ ($e$ is the negative electron charge)$,$ $f^{\left( \mp \right) }=$
$f_{\mathrm{F}}\left( \varepsilon -eV\,\Theta (-z)\right) .$ Figure 3
illustrates the occupied energy states in the two conductors for both signs
of the applied bias $eV$. At $eV>0$ the electrons flow from the bulk
conductor (the tip) into the film and, vice-versa, at $eV<0$ they flow from
the film into the massive conductor. 
\begin{figure}[tbp]
\includegraphics[width=15cm,angle=0]{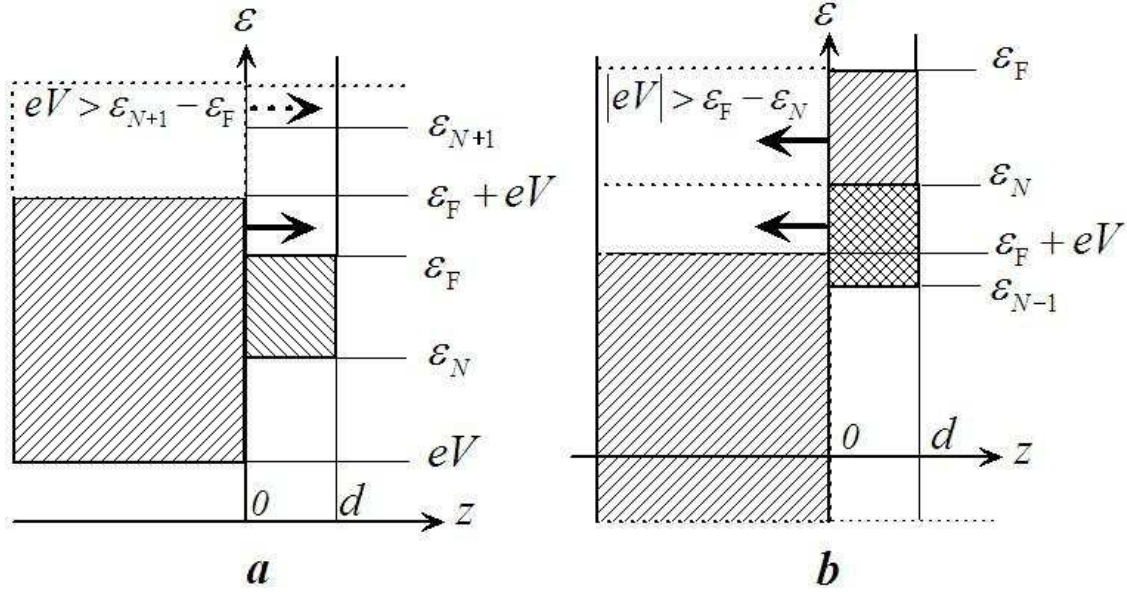}
\caption{Illustration of the occupied energy states at zero temperature in
the two conductors for both signs of the applied bias $eV$: (a) $eV>0$, (b) $%
eV<0$.}
\label{fig3}
\end{figure}
The total current through the area of the contact can be found by
integration over the flux $J^{\left( \pm \right) }$ in both directions 
\begin{gather}
{I(V)=\frac{1}{2\pi d}\int_{-\infty }^{\infty }d\mathbf{\kappa }%
\sum\limits_{n=1}^{\infty }J^{\left( -\right) }f_{\mathrm{F}}\left(
\varepsilon \right) \left( 1-f_{\mathrm{F}}\left( \varepsilon -eV\right)
\right) -}  \label{I(V)} \\
{-\frac{2}{\left( 2\pi \right) ^{3}}\int_{-\infty }^{\infty }d\mathbf{k}J{%
^{\left( +\right) }f_{\mathrm{F}}\left( \varepsilon -eV\right) \left( {%
\newline
1-f_{\mathrm{F}}\left( \varepsilon \right) }\right) },}  \notag
\end{gather}%
In Eq.~(\ref{I(V)}) we integrate over the wave vector $\mathbf{k}$ in the
semi-infinite conductor for the current in the negative direction (first
term), and integrate over the two-dimensional wave vector \textbf{$\kappa $}
and sum over the discrete quantum number $n$ for the opposite direction of
the current (second term).

For simplicity we will take the temperature to be zero. In this case the
electric current is defined by electrons passing the contact in one
direction only, depending on the sign of the applied bias. The flux $%
J^{\left( \pm \right) }$ integrated over the area of the contact is
calculated in the usual way 
\begin{equation}
J^{\left( \pm \right) }=\frac{\left\vert e\right\vert \hbar }{m^{\ast }}%
\int\limits_{0}^{a}d\rho \rho \int\limits_{0}^{2\pi }d\phi \func{Im}\left[
\psi _{1}^{\ast }\left( \mathbf{\rho },z\right) \frac{\partial }{\partial z}%
\psi _{1}\left( \mathbf{\rho },z\right) \right] _{z=\pm 0},  \label{J}
\end{equation}%
where $\mathbf{\rho =}\left( \rho \cos \phi ,\rho \sin \phi \right) .$ The
wave function $\psi _{1}\left( \mathbf{\rho },z\right) $ should be taken as
the wave transmitted through the contact, given by equations (AI.5) and
(AII.6) with $\Bbbk =k_{z}$ for electron flux from the tip to the sheet, $%
J^{\left( +\right) }$, and by equations (AI.6) and (AII.7) with $\Bbbk
=k_{zn}$ $\left( n=1,2...\right) $ for fluxes $J^{\left( -\right) }$ in the
opposite direction. The energy shift $eV$ in the region $z<0$ should be
taken into account, which for our choice of the reference point of energy
(see Fig.~\ref{fig3}) implies that the absolute value of the electron wave
vector in the half-space $z<0$ is given by $\widetilde{k}=\sqrt{2m^{\ast
}\left( \varepsilon -eV\right) }/\hbar .$

For the tunnel point contact (\textit{tpc}) the flux can be expressed in
terms of the wave function in the contact plane (AI.1), and we obtain, 
\begin{equation}
J_{tpc}^{\left( +\right) }\simeq \frac{\pi ^{4}\left\vert e\right\vert
a^{4}\hbar ^{5}\widetilde{k}^{2}\cos ^{2}\vartheta }{12m^{\ast
3}d^{3}U_{0}^{2}}N\left( N+1\right) \left( 2N+1\right) ,  \label{Jtpc(+)}
\end{equation}%
and 
\begin{equation}
J_{tpc}^{\left( -\right) }\simeq -\frac{\pi \left\vert e\right\vert
a^{4}\hbar ^{5}\widetilde{k}^{3}k_{zn}^{2}}{6m^{\ast 3}U_{0}^{2}}.
\label{Jtpc(-)}
\end{equation}%
Here, $\vartheta $ is the angle between the vector $\mathbf{k}$ and the $z$
axis, and $N\left( k\right) =[kd/\pi ]$ with $[x]$ the integer part of $x.$

\begin{figure}[tbp]
\includegraphics[width=10cm,angle=0]{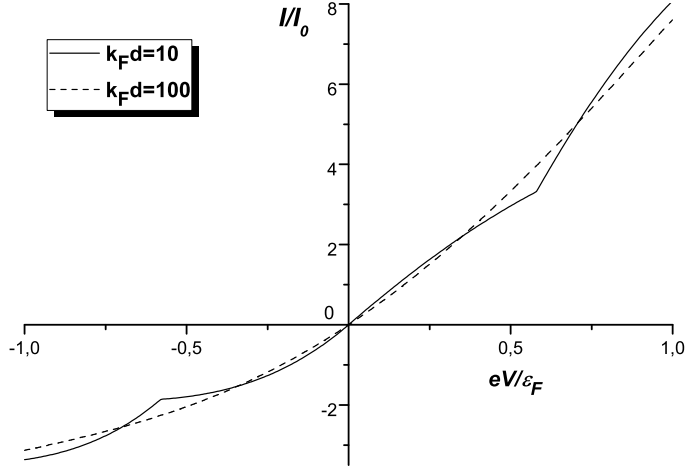}
\caption{Dependence of the total current, $I\left( V\right) $, on the
applied bias over the point contact for two choices of the thicknesses of
the metal film. The constant $I_{0}$ is given by Eq.~(\protect\ref{Itpc1})
or Eq.~(\protect\ref{Impc1}). }
\label{fig-I}
\end{figure}

For a metallic point contact (\textit{mpc}) without barrier the expressions
for the flux $J_{mpc}^{\left( \pm \right) }$ are written by means of
Eqs.~(AII.4), (AII.9)-(AII.11), 
\begin{equation}
J_{mpc}^{\left( +\right) }\simeq \frac{\pi ^{2}\left\vert e\right\vert \hbar
a^{6}\widetilde{k}^{2}\cos ^{2}\vartheta }{9m^{\ast }d^{3}}N\left(
N+1\right) \left( 2N+1\right) ,  \label{Jor(+)}
\end{equation}%
and%
\begin{equation}
J_{mpc}^{\left( -\right) }\simeq -\frac{8\pi \left\vert e\right\vert \hbar
a^{6}\widetilde{k}^{3}k_{zn}^{2}}{9m^{\ast }}  \label{Jor(-)}
\end{equation}%
Substituting Eqs.~(\ref{Jtpc(+)})-(\ref{Jor(-)}) into the general expression
(\ref{I(V)}) we find the current-voltage characteristic of the system 
\begin{equation}
I\left( V\right) =\frac{I_{0}}{\left( k_{\mathrm{F}}d\right) ^{3}}%
\int\limits_{k_{\mathrm{F}}}^{\widetilde{k}_{\mathrm{F}}}\frac{dkk^{2}}{k_{%
\mathrm{F}}^{5}}(k^{2}-\frac{2meV}{\hbar ^{2}})S_{2}\left( k\right) ;\quad
eV\geqslant 0  \label{I(eV>0)}
\end{equation}%
and 
\begin{gather}
I\left( V\right) =-\frac{I_{0}}{\left( k_{\mathrm{F}}d\right) ^{3}}\left\{
S_{2}\left( k_{\mathrm{F}}\right) \left( \frac{1}{5}+\frac{2}{3}\frac{%
\left\vert eV\right\vert }{\varepsilon _{\mathrm{F}}}+\frac{1}{3}\left( 
\frac{\left\vert eV\right\vert }{\varepsilon _{\mathrm{F}}}\right)
^{2}\right) \right. +  \label{I(eV<0)} \\
\left[ S_{3}\left( \widetilde{k}_{\mathrm{F}}\right) -S_{3}\left( k_{\mathrm{%
F}}\right) \right] \frac{\pi }{k_{\mathrm{F}}d}\left( \frac{\left\vert
eV\right\vert }{\varepsilon _{\mathrm{F}}}\right) ^{2}+  \notag \\
\quad \frac{2}{3}\left[ S_{5}\left( \widetilde{k}_{\mathrm{F}}\right)
-S_{5}\left( k_{\mathrm{F}}\right) \right] \left( \frac{\pi }{k_{\mathrm{F}}d%
}\right) ^{3}\frac{\left\vert eV\right\vert }{\varepsilon _{\mathrm{F}}}+%
\frac{1}{5}\left[ S_{7}\left( \widetilde{k}_{\mathrm{F}}\right) -S_{7}\left(
k_{\mathrm{F}}\right) \right] \left( \frac{\pi }{k_{\mathrm{F}}d}\right)
^{5}-  \notag \\
S_{2}\left( \widetilde{k}_{\mathrm{F}}\right) \frac{\widetilde{k}_{\mathrm{F}%
}}{5k_{\mathrm{F}}}\left( 1+\frac{4}{3}\frac{\left\vert eV\right\vert }{%
\varepsilon _{\mathrm{F}}}-\frac{8}{3}\left( \frac{\left\vert eV\right\vert 
}{\varepsilon _{\mathrm{F}}}\right) ^{2}\right) ;\quad eV\leqslant 0  \notag
\end{gather}%
where $\varepsilon _{\mathrm{F}}=\hbar ^{2}k_{\mathrm{F}}^{2}/2m^{\ast }$ is
the Fermi energy, 
\begin{equation}
I_{0,tpc}=\frac{\left\vert e\right\vert \pi ^{2}a^{4}\hbar ^{5}k_{\mathrm{F}%
}^{8}}{12m^{\ast 3}U_{0}^{2}},  \label{Itpc1}
\end{equation}

\begin{equation}
I_{0,mpc}=\frac{e\hbar a^{6}k_{\mathrm{F}}^{8}}{9m^{\ast }},  \label{Impc1}
\end{equation}%
$S_{m}\left( k\right) $ is a finite sum of $m-th$ powers of integer numbers, 
\begin{equation}
S_{m}\left( k\right) =\sum\limits_{n=1}^{N\left( k\right) }n^{m}.  \label{Sm}
\end{equation}%
Note that $S_{m}\left( k\right) \equiv $ $H_{-m}\left( N\right) ,$ where $%
H_{m}\left( n\right) $ are generalized harmonic numbers. The current is
plotted in Fig.~\ref{fig-I} as a function of bias voltage for two choices of
the film thickness. 
Differentiating Eqs. (\ref{I(eV>0)}) and (\ref{I(eV<0)}) with respect to
voltage we obtain the differential conductance $G\left( V\right) =dI/dV$ for
a point contact with radius $a\ll \lambdabar _{\mathrm{F}}$, 
\begin{equation}
G\left( V\right) =G_{1}\left\{ \frac{\widetilde{k}_{\mathrm{F}}}{2k_{\mathrm{%
F}}}S_{2}\left( \widetilde{k}_{\mathrm{F}}\right) -\frac{1}{k_{\mathrm{F}%
}^{3}}\int\limits_{k_{\mathrm{F}}}^{\widetilde{k}_{\mathrm{F}%
}}dkk^{2}S_{2}\left( k\right) \right\} ,\qquad \qquad eV\geqslant 0;
\label{G(eV>0)}
\end{equation}%
\begin{gather}
G\left( V\right) =G_{1}\left\{ \frac{4}{3}\left[ 1+\frac{\left\vert
eV\right\vert }{\varepsilon _{\mathrm{F}}}\right] S_{2}\left( k_{\mathrm{F}%
}\right) +4\frac{\left\vert eV\right\vert }{\varepsilon _{\mathrm{F}}}\frac{%
\pi }{k_{\mathrm{F}}d}\left[ S_{3}\left( \widetilde{k}_{\mathrm{F}}\right)
-S_{3}\left( k_{\mathrm{F}}\right) \right] \right. +  \label{G(eV<0)} \\
\left. \frac{4}{3}\left( \frac{\pi }{k_{\mathrm{F}}d}\right) ^{3}\left[
S_{5}\left( \widetilde{k}_{\mathrm{F}}\right) -S_{5}\left( k_{\mathrm{F}%
}\right) \right] -\frac{k_{\mathrm{F}}}{3\widetilde{k}_{\mathrm{F}}}\left[
1+4\frac{\left\vert eV\right\vert }{\varepsilon _{\mathrm{F}}}-8\left( \frac{%
\left\vert eV\right\vert }{\varepsilon _{\mathrm{F}}}\right) ^{2}\right]
S_{2}\left( \widetilde{k}_{\mathrm{F}}\right) \right\} ,\quad eV\leqslant 0.
\notag
\end{gather}%
In the limit $eV\rightarrow 0$ the zero-bias conductance taken from both
sides coincides, as it should, 
\begin{equation}
G\left( 0\right) =G_{1}S_{2}\left( k_{\mathrm{F}}\right) =\frac{G_{1}}{6}N_{%
\mathrm{F}}\left( N_{\mathrm{F}}+1\right) \left( 2N_{\mathrm{F}}+1\right) ,
\label{G(0)}
\end{equation}%
where $N_{\mathrm{F}}=N\left( k_{\mathrm{F}}\right) $, and $G_{1}$ is the
conductance of the contact between the bulk conductor (the tip) and a thin
film that has only a single energy level available below $\varepsilon _{%
\mathrm{F}}$ for the motion along $z$, 
\begin{equation}
G_{1}=G_{0}\left( 0\right) \frac{3\pi ^{3}}{\left( k_{\mathrm{F}}d\right)
^{3}}.
\end{equation}%
$G_{0}\left( 0\right) $ is the conductance of a contact between two
conducting unbound half-spaces. For a tunnel point contact this is given by 
\cite{Kulik,AKR}%
\begin{equation}
G_{0,tpc}\left( 0\right) =\left( \frac{k_{\mathrm{F}}\hbar ^{2}}{m^{\ast
}U_{0}}\right) ^{2}\frac{e^{2}\left( k_{\mathrm{F}}a\right) ^{4}}{36\pi
\hbar },  \label{G0tpc}
\end{equation}%
and for a metallic point contact we have \cite{ISh}, 
\begin{equation}
G_{0,mpc}\left( 0\right) =\frac{8e^{2}\left( k_{\mathrm{F}}a\right) ^{6}}{%
27\pi ^{3}\hbar }.  \label{G0or}
\end{equation}%
For $d\rightarrow \infty $ Eqs.~(\ref{G(eV>0)}) and (\ref{G(eV<0)})
transform into the known voltage dependence of the conductance for a point
contact between unbound conducting half-spaces \cite{Avotina08mi}, 
\begin{equation}
G_{0}\left( V\right) =G_{0}\left( 0\right) \left[ 1+\frac{\left\vert
eV\right\vert }{\varepsilon _{\mathrm{F}}}-\frac{1}{3}\left( \frac{%
\left\vert eV\right\vert }{\varepsilon _{\mathrm{F}}}\right) ^{3}\right] .
\label{G0}
\end{equation}

\begin{figure}[tbp]
\includegraphics[width=10cm,angle=0]{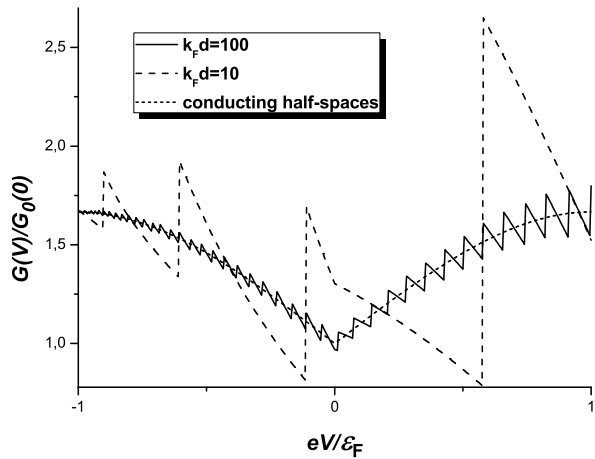}
\caption{Dependence of the normalized differential conductance, $G\left(
V\right) /G_{0}\left( 0\right) $, on the applied bias over the point contact
for two choices of the thicknesses of the metal film. The voltage dependence
for a point contact between two semi-infinite bulk conductors is shown for
comparison (short-dashed curve).}
\label{G}
\end{figure}
The dependence of the differential conductance $G\left( V\right) $ for both
signs of applied voltage is illustrated in Fig.~\ref{G}. For comparison the
dependence $G_{0}\left( V\right) /G_{0}\left( 0\right) $ from Eq.~(\ref{G0})
is also shown.

\section{Discussion}

Thus, in the framework of the model illustrated in Fig.~\ref{fig1} we have
obtained the current-voltage characteristic and the differential conductance
for a contact on the surface of a thin metal film. Under the assumption that
the contact radius $a$ is much smaller than the Fermi wave length $%
\lambdabar _{\mathrm{F}}$ we found asymptotically exact formulas for the
dependence of the total current $I\left( V\right) $ (Eqs. (\ref{I(eV>0)})
and (\ref{I(eV<0)})) and the contact conductance $G\left( V\right) $ (Eqs.~(%
\ref{G(eV>0)}),~(\ref{G(eV<0)})) on the applied voltage. In the limit of
zero temperature and neglecting scattering processes we have demonstrated
that the $I\left( V\right) $ dependence has kinks and $G\left( V\right) $
undergoes jumps at the same values of applied bias $eV$ (see Fig.\ref{fig-I}
and Fig.\ref{G}) These events result from the size quantization of the
electron spectrum in the film.

The results obtained show that even in Ohm's-law approximation (\ref{G(0)}), 
$eV\rightarrow 0$, the conductance $G\left( V\right) $ is not simply
proportional to the electron density of states (DOS) in the isolated film,%
\begin{equation}
\rho _{f}\left( \varepsilon \right) =\frac{m^{\ast }N_{F}}{\pi \hbar ^{2}d}.
\label{dos}
\end{equation}%
It is remarkable that the dependence of the conductance $G\left( 0\right) $ (%
\ref{G(0)}) on the number of quantum levels $N_{\mathrm{F}}$ is the same
for, both, tunnel and metallic point contacts. This fact shows that such
dependence is not sensitive to the model taken for the potential barrier,
and that it is the result of the point-contact geometry. Recently, the
relationship between the differential conductance and the local density of
states has been studied in a tight-binding approximation for tunnel
junctions, where the junction geometry can be varied between the limiting
cases of a point-contact and a planar junction \cite{Berthod}. In the
framework of a real-space Keldysh formalism the authors of Ref.~\cite%
{Berthod} have shown that the differential conductance is not, in general,
proportional to the sample DOS for planar junctions, although features of
the DOS may be present.

From Eqs.~(\ref{G(eV>0)}) and (\ref{G(eV<0)}) it follows that the the
conductance is non-symmetric in the applied bias. This asymmetry can be
explained as follows: Let $eV>0$ and electrons tunnel from the bulk
conductor into the film (Fig.~\ref{fig3}a) in which $N_{\mathrm{F}}$
subbands of the size quantization are partially filled. If the bias $eV$ is
smaller than the distance $\Delta \varepsilon $ between the Fermi level $%
\varepsilon _{\mathrm{F}}$ and the bottom of the next (empty) subband $%
\varepsilon _{N+1}=\pi ^{2}\hbar ^{2}\left( N_{\mathrm{F}}+1\right)
^{2}/2m^{\ast }d^{2},$ $\Delta \varepsilon =\varepsilon _{N+1}-\varepsilon _{%
\mathrm{F}},$ the electron can tunnel into any of the $N_{\mathrm{F}}$
subbands. At $eV=\Delta \varepsilon $ tunneling into the $\left( N_{\mathrm{F%
}}+1\right) $-th subband becomes possible and the conductance $G\left(
V\right) $ undergoes a positive jump. Such jumps are repeated for increasing
voltage for all higher subbands. For $eV<0,$ when electrons tunnel from the
thin film into bulk metal (Fig.~\ref{fig3}b) the situation is somewhat
different. If the bias $\left\vert eV\right\vert $ becomes larger than
distance $\Delta \varepsilon $ between the bottom of the last partially
filled subband $\varepsilon _{N}=\pi ^{2}\hbar ^{2}N_{\mathrm{F}%
}^{2}/2m^{\ast }d^{2}$ and Fermi energy, $\Delta \varepsilon =\varepsilon _{%
\mathrm{F}}-\varepsilon _{N}$, the contribution of the $N_{\mathrm{F}}$-th
subband to the tunnel current does not depend on the voltage because for any 
$\left\vert eV\right\vert >\Delta \varepsilon $ all electrons of this
subband can tunnel into the bulk states of the left conductor. For this
reason the differential conductance drops for values of $\left\vert
eV\right\vert $ coinciding with bottoms of subbands of size quantization in
the film. The distance between neighboring jumps of the conductance on the
voltage scale equals the distance between energy levels $\Delta \varepsilon
_{N}=\varepsilon _{N+1}-\varepsilon _{N}=\pi ^{2}\hbar ^{2}\left( 2N_{%
\mathrm{F}}+1\right) /2m^{\ast }d^{2}.$ For $eV<0$ the number of conductance
jumps is finite and equals the number of discrete levels below Fermi surface 
$N_{\mathrm{F}}.$ The asymmetry around $V=0$ and the general shape of the
jumps in the conductance can be recognized in the experiments, see e.g. \cite%
{Jiang}. In the special case of a 2D electron system, which has only one
level in the potential well, there is a single negative jump of $G\left(
V\right) .$ Such a jump has been observed in Ref.~\cite{Burgi} by STM
investigations of the 2D electron gas at noble-metal surfaces. For $eV>0$
the number of conductance jumps formally is not restricted. However, for $%
eV>\varepsilon _{\mathrm{F}}$ our approach is no longer applicable and the
influence of field emission on the tunnel current must be taken into account 
\cite{Kempen,Mongenstern}

The observation of manifestations of the size quantization in the STM
conductance requires a few conditions which must be fulfilled: The distance
between the energy levels must be large enough and should satisfy the
condition $\Delta \varepsilon _{N}\gg \hbar /\tau ,T,$ where $\tau $ is the
mean scattering time of the electrons in the film and $T$ is the
temperature. The surfaces of the metal film in the region of the contact
must be atomically smooth \cite{Tavger}. When a finite lifetime of the
quantized states becomes relevant, the temperature broadening of the Fermi
function, or surface imperfections need to be taken into account this will
result in a rounding of the jumps in the curve $G\left( V\right) $ presented
in Fig.~\ref{G} (Eqs.\ref{G(eV>0)}, \ref{G(eV<0)}), which was plotted under
assumptions of perfectly specular surfaces, $T=0,$ and $\tau \rightarrow
\infty .$ With these restrictions taken into account the current-voltage
curves in Fig.~\ref{fig-I} give a fair qualitative description of the
experimental results of Ref.~\cite{Alt2}.

It can be easily seen that the results obtained have a more wide domain of
applicability than that of a rectangular well for the conducting film. For
any model of the potential which restricts the electron motion in one
direction the differential conductance has a step-like dependence on the
applied bias with distances between the steps equal to the distances between
the quantum levels.

\section{Conclusion.}

In summary, we have investigated the conductance of ultra small contacts,
for which the radius is smaller than the Fermi wave length, on top of the
surface of a thin film. The discreteness of the component of the electron
momentum transverse to the film surface is taken into account, where the
distance between the electron energy levels due to the size quantization is
assumed to be larger than the temperature. Both, a contact with a potential
barrier of low transparency, and a contact without barrier have been
considered. In framework of our model, using a $\delta $-function potential
barrier, the current-voltage characteristic $I\left( V\right) $ of the
system and differential conductance $G\left( V\right) $ have been obtained.
We predict a sawtooth dependence of $G\left( V\right) $ on the applied bias
and show that the distance between neighboring jumps is equal to the
distance between neighboring energy levels of size quantization, i.e. this
dependence can be used for spectroscopy of size quantization levels. At $%
eV>0 $ the jumps in the conductance are positive and correspond to distances
between levels above the Fermi surface, while $G\left( V\right) $ undergoes
negative jumps for $eV<0$, the distances between which are equal to the
distances between the levels below the Fermi surface. The predicted
quantization of the conductance can be observed in STS experiments, and the
shape of the theoretical curves agrees well with experiments.

\section{Appendix I: Electron tunneling between the tip and the thin film.}

We search a solution to Eq.~(\ref{Schrod}) at $V=0$ in the form of a sum $%
\psi =\psi _{0}+\psi _{1}$ for the incident and backscattered waves, and $%
\psi =\psi _{1}$ for the transmitted wave. Here, $\psi _{0}$, as given by
Eqs.~(\ref{psi0z<0}), (\ref{psi0z>0}), is the unperturbed wave function that
does not depend on the barrier amplitude $U_{0}$, while $\psi _{1}\sim
1/U_{0}$ gives the first order correction. Substituting the wave function
into the boundary conditions (\ref{z=0}) and (\ref{jump}) one should match
terms of the same order in $1/U_{0}.$ As a result the boundary condition (%
\ref{jump}) becomes \cite{Kulik}%
\begin{equation}
\psi _{1}\left( \mathbf{\rho },0\right) =-\frac{i\Bbbk \hbar ^{2}}{m^{\ast
}U_{0}}e^{-i\mathbf{\kappa \rho }}\Theta \left( a-\rho \right) ,  \tag{AI.1}
\end{equation}%
where $\Bbbk =k_{z},$ when the wave is incident to the contact from the tip
side, and $\Bbbk =k_{zn}$ when the wave arrives at the contact from the
sheet. For $ka\ll 1$ we have in the plane of the contact \textbf{$\kappa $}$%
\mathbf{\rho }\ll 1$ and we can neglect the exponent in the boundary
condition (AI.1).

The function $\psi _{1}\left( \mathbf{\rho },z\right) $ can be represented
as a Fourier integral%
\begin{equation}
\psi _{1}\left( \mathbf{\rho },z\right) =\int\limits_{-\infty }^{\infty }d%
\mathbf{\kappa }^{\prime }e^{-i\mathbf{\kappa }^{\prime }\mathbf{\rho }}\Psi
\left( \mathbf{\kappa }^{\prime },z\right) .  \tag{AI.2}
\end{equation}%
The Fourier components in (AI.2) should satisfy the zero boundary condition
at $z=d$, but are otherwise freely propagating along $z$, 
\begin{equation}
\Psi \left( \mathbf{\kappa }^{\prime },z\right) =\Psi \left( \mathbf{\kappa }%
^{\prime },0\right) \frac{\sin k_{z}^{\prime }\left( z-d\right) }{\sin
k_{z}^{\prime }d},\qquad \qquad \qquad 0\leqslant z\leqslant d,  \tag{AI.3}
\end{equation}%
\begin{equation}
\Psi \left( \mathbf{\kappa }^{\prime },z\right) =\Psi \left( \mathbf{\kappa }%
^{\prime },0\right) \exp \left( -ik_{z}^{\prime }z\right) ,\qquad \qquad
\qquad z\leqslant 0,  \tag{AI.4}
\end{equation}%
with $k_{z}^{\prime }=\sqrt{k^{2}-\kappa ^{\prime 2}},$ $k=\sqrt{2m^{\ast
}\varepsilon }/\hbar .$ From Eqs.~(AI.1), (AI.2) it follows that%
\begin{equation}
\Psi \left( \mathbf{\kappa }^{\prime },0\right) =\frac{1}{(2\pi )^{2}}%
\int\limits_{-\infty }^{\infty }d\mathbf{\rho }e^{i\mathbf{\kappa }^{\prime }%
\mathbf{\rho }}\psi \left( \mathbf{\rho },0\right) =-\frac{i\Bbbk \hbar ^{2}a%
}{2\pi m^{\ast }U_{0}}\frac{J_{1}\left( \kappa ^{\prime }a\right) }{\kappa
^{\prime }}.  \tag{AI.4}
\end{equation}%
Substituting this into Eq.~(AI.2) we find the wave functions for the
electrons transmitted through the contact as, 
\begin{equation}
\psi _{1}\left( \mathbf{\rho },z\right) =\frac{i\Bbbk \hslash ^{2}a}{m^{\ast
}U_{0}}\int\limits_{0}^{\infty }d\kappa ^{\prime }J_{0}\left( \kappa
^{\prime }\rho \right) J_{1}\left( \kappa ^{\prime }a\right) \frac{\sin
k_{z}^{\prime }\left( d-z\right) }{\sin k_{z}^{\prime }d},\qquad
0<z\leqslant d,  \tag{AI.5}
\end{equation}%
\begin{equation}
\psi _{1}\left( \mathbf{\rho },z\right) =\frac{i\Bbbk \hslash ^{2}a}{m^{\ast
}U_{0}}\int\limits_{0}^{\infty }d\kappa ^{\prime }J_{0}\left( \kappa
^{\prime }\rho \right) J_{1}\left( \kappa ^{\prime }a\right) \exp \left(
-ik_{z}^{\prime }z\right) ,\qquad \qquad z<0  \tag{AI.6}
\end{equation}%
where $J_{n}\left( x\right) $ is the Bessel function of the first kind.

\section{Appendix II: metallic point contact between STM tip and metal film.}

Here we consider a point contact without potential barrier in the plane of
the interface. When the contact radius is small, $ka\ll 1$, we can use
perturbation theory for the electron wave function in the limit $%
a\rightarrow 0.$ In zeroth approximation the wave functions are given by
Eqs.~(\ref{psi0z<0}) and (\ref{psi0z>0}). The first order correction, $\psi
_{1}\left( \mathbf{\rho },0\right) $, to the wave function in the plane of
the contact can be found by the method proposed in \cite{ISh}. For distances 
$r\ll \lambda $ from the contact we can neglect the second term in the Schr%
\"{o}dinger equation (\ref{Schrod}) and it reduces to the Laplace equation.
We express the wave function in coordinates of an oblate ellipsoid of
revolution $\left( \sigma ,\tau ,\varphi \right) $, with $\sigma \geqslant 0$
and $-1\leqslant \tau \leqslant 1$. As a consequence of the cylindrical
symmetry of the problem the wave function $\psi _{1}\left( \sigma ,\tau
\right) $ does not depend on $\varphi .$ The interface corresponds to $\tau
=0$ and the plane of the orifice is at $\sigma =0.$ In these coordinates we
obtain the equation%
\begin{equation}
\frac{\partial }{\partial \sigma }\left[ \left( 1+\sigma ^{2}\right) \frac{%
\partial \psi _{1}}{\partial \sigma }\right] +\frac{\partial }{\partial \tau 
}\left[ \left( 1-\tau ^{2}\right) \frac{\partial \psi _{1}}{\partial \tau }%
\right] =0,  \tag{AII.1}
\end{equation}%
with the boundary condition at the interface%
\begin{equation}
\psi _{1}\left( \sigma >0,\tau =0\right) =0.  \tag{AII.2}
\end{equation}%
The solution of the boundary problem (AII.1), (AII.2) is%
\begin{equation}
\psi _{1}\left( \sigma ,\tau \right) =\tau \left[ c_{1}\sigma +c_{2}\left(
1+\sigma \arctan \sigma \right) \right] ,  \tag{AII.3}
\end{equation}%
where $c_{1}$ and $c_{2}$ are constants. For $\sigma =0$ Eq.~(AII.3) gives
the function $\psi _{1}\left( \mathbf{\rho },z\right) $ in the plane of the
contact $z=0,$ $\rho \leqslant a$%
\begin{equation}
\psi _{1}\left( \mathbf{\rho },0\right) =c_{2}\sqrt{1-\frac{\rho ^{2}}{a^{2}}%
}.  \tag{AII.4}  \label{AII.4}
\end{equation}%
As in Appendix I, we express $\psi _{1}\left( \mathbf{\rho },z\right) $ as a
Fourier integral and, using the Eq.~(AII.4), we find for the Fourier
components, 
\begin{equation}
\Psi \left( \mathbf{\kappa }^{\prime },0\right) =\frac{1}{(2\pi )^{2}}%
\int\limits_{-\infty }^{\infty }d\mathbf{\rho }e^{i\mathbf{\kappa }^{\prime }%
\mathbf{\rho }}\psi _{1}\left( \mathbf{\rho },0\right) =c_{2}a\frac{%
j_{1}\left( \kappa ^{\prime }a\right) }{\kappa ^{\prime }},  \tag{AII.5}
\end{equation}%
where $j_{1}\left( x\right) $ is the spherical Bessel function of the first
kind. Substituting Eq.~(AII.5) into Eq.~(AI.2) and using Eqs.~(AI.3), (AI.4)
we obtain 
\begin{equation}
\psi _{1}\left( \rho ,z\right) =\frac{c_{2}a}{2\pi }\int\limits_{0}^{\infty
}d\kappa ^{\prime }J_{0}\left( \kappa ^{\prime }\rho \right) j_{1}\left(
\kappa ^{\prime }a\right) \frac{\sin k_{z}^{\prime }\left( d-z\right) }{\sin
k_{z}^{\prime }d},\quad 0<z\leqslant d,  \tag{AII.6}
\end{equation}%
and%
\begin{equation}
\psi _{1}\left( \rho ,z\right) =\frac{c_{2}a}{2\pi }\int\limits_{0}^{\infty
}d\kappa ^{\prime }J_{0}\left( \kappa ^{\prime }\rho \right) j_{1}\left(
\kappa ^{\prime }a\right) e^{-ik_{z}^{\prime }z},\qquad \qquad \qquad z<0. 
\tag{AII.7}
\end{equation}%
The constant $c_{2}$ must be found from the boundary condition (\ref{jump})
at $U_{0}=0$, which for this case takes the form%
\begin{equation}
\frac{\partial \psi _{1}\left( \rho ,+0\right) }{\partial z}-\frac{\partial
\psi _{1}\left( \rho ,-0\right) }{\partial z}-2i\Bbbk =0.  \tag{AII.8}
\end{equation}%
The meaning of the symbol $\Bbbk $ is explained below Eq.(AI.1).
Differentiating Eqs.~(AII.6) and (AII.7) with respect to $z$ and calculating
the integrals in the limit of small $a$ we find, 
\begin{equation}
\left. \frac{\partial \psi _{1}}{\partial z}\right\vert _{z=+0}\simeq \frac{%
c_{2}a}{2\pi }\left( -\frac{\pi }{2a^{2}}+i\frac{\pi ^{3}a}{18d^{3}}%
N(N+1)(2N+1)\right) ,  \tag{AII.9}  \label{dpsiz+}
\end{equation}%
\begin{equation}
\left. \frac{\partial \psi _{1}}{\partial z}\right\vert _{z=-0}\simeq \frac{%
c_{2}a}{2\pi }\left( \frac{\pi }{2a^{2}}+i\frac{\pi k^{3}a}{9}\right) , 
\tag{AII.10}  \label{dpsiz-}
\end{equation}%
where $N=[kd/\pi ]$ with $[x]$ the integer part of $x.$ Substituting
Eqs.~(AII.9) and (AII.10) into (AII.8) in leading approximation in $a,$ in
which only first terms in brackets (proportional to $1/a^{2}$) should be
taken into account, we find for the unknown constant, 
\begin{equation}
c_{2}\simeq 2i\Bbbk a.  \tag{AII.11}  \label{c2}
\end{equation}

\end{document}